\documentclass[twocolumn,english,aps,english,preprintnumbers,groupedaddress,amsmath,amssymb,pra]{revtex4}
\usepackage[T1]{fontenc}
\usepackage[latin9]{inputenc}
\usepackage{amsmath}
\usepackage{amssymb}
\usepackage{graphicx}
\usepackage{esint}

\makeatletter
\@ifundefined{textcolor}{}
{%
 \definecolor{BLACK}{gray}{0}
 \definecolor{WHITE}{gray}{1}
 \definecolor{RED}{rgb}{1,0,0}
 \definecolor{GREEN}{rgb}{0,1,0}
 \definecolor{BLUE}{rgb}{0,0,1}
 \definecolor{CYAN}{cmyk}{1,0,0,0}
 \definecolor{MAGENTA}{cmyk}{0,1,0,0}
 \definecolor{YELLOW}{cmyk}{0,0,1,0}
}

\makeatother

\usepackage{babel}
\begin{document}

\title{On-resonance material fast light }

\author{Bruno Macke}

\author{Bernard S\'{e}gard}
\email{bernard.segard@univ-lille.fr}

\affiliation{Laboratoire de Physique des Lasers, Atomes et Mol\'{e}cules , CNRS et
Universit\'{e} de Lille, 59655 Villeneuve d'Ascq, France}

\date{\today}
\begin{abstract}
We theoretically revisit the problem of the propagation of coherent
light pulses through a linear medium when the carrier frequency of
the pulses coincides with the minimum of a narrow dip in the medium
transmission. Considering realistic contrasts between the maximum
and minimum transmission of the medium and incident pulses of strictly
finite duration, we combine temporal and spectral approaches to obtain
analytical expressions of the transmitted pulse that reproduce the
main features of the exact numerical solutions derived by fast Fourier
transform. A special attention is paid to the advance of the pulse
maximum over that of a pulse covering the same distance in vacuum
and the ratio of this advance to the pulse duration (fractional advance)
is optimized.
\end{abstract}
\maketitle

\section{Introduction\label{sec:Introduction}}

Fast light currently refers to situations in which a smooth pulse
of coherent light is transmitted by a linear optical system in such
a manner that the maximum of the output pulse anticipates that of
a pulse having covered the same distance in vacuum. As extensively
discussed in the literature (see, e.g., \cite{chi97,bo02}), the phenomenon
is not at odds with the relativistic causality. It appears in particular
that the maximum of the output pulse is not as a direct reflection
of that of the input pulse but results from the action of the system
on the early part of the latter. A direct experimental evidence of
this point is reported in \cite{to14}. Fast light can be obtained
when the system transmission displays a well-marked, narrow dip at
the carrier frequency of the input pulse \cite{ma05}. Note however
that this condition is not always sufficient. For optical systems
that are not minimum-phase-shift \cite{pa87}, in particular those
involving mirrors \cite{wa02a,ma16a} or polarizers \cite{so03,bru04,ma16b},
it may even occur that a same transmission profile leads to either
fast or slow light. In the case of purely propagative systems as considered
in the following, fast light simply originates in the dispersive properties
of the medium (material fast light) and the Kramers-Kronig relations
apply in their common form. Even in this case a dip in the transmission
profile does not always entail fast light. For instance when the dip
corresponds to a minimum of transmission between two gain lines, there
is a range of parameters for which there is no fast light whatever
the dip depth is \cite{ma03}. 

The simplest way to observe material fast light is to exploit the
anomalous dispersion associated with an isolated absorption line of
a passive medium. Experiments have been performed on various materials
including semi-conductor \cite{chu82}, molecular gas \cite{se85},
atomic vapour at room temperature \cite{ta03}, hot atomic vapour
\cite{ke12} and clouds of cold atoms \cite{je16}, with propagation
distances going from 390 nanometers \cite{ke12} to 24 meters \cite{se85}.
In order to control the pulse advance, the transmission dip is artificially
created in other experiments by applying supplementary fields interacting
non linearly with the medium and taking advantage of the phenomena
of electromagnetically induced absorption \cite{ak02,go02,ki03,ka04,mi04,br08,ak10},
Raman effect \cite{wa00,ste03a}, Brillouin effect \cite{chi07} and
four wave mixing \cite{gla12,sw17}.

In most experimental reports on material fast light, the emphasis
is made on the absolute value of the advance of the pulse maximum
over that of a pulse propagating in vacuum and on the corresponding
value of the velocity often confused with the group velocity. In fact
the true experimental challenge is to obtain advances which are significant
compared to the pulse duration with acceptable pulse distortion. On
the other hand, in the numerous theoretical papers on the subject,
the proof of fast light is generally reduced to that of a superluminal
group velocity without simulation evidencing a visible advance of
the pulse maximum.

The present article is theoretical but devotes a special attention\emph{
to actual or feasible experiments}. Its purpose is to clarify some
points about material fast light and to examine what are the practical
limitations to the phenomenon, Mother Nature resisting to a violation
of her principles even when this violation is only apparent. To do
so, we come back to the basic system consisting in a passive medium
with an isolated absorption line \cite{chu82,se85,ta03,ke12,je16}.
The simplicity of this system, favourable from an experimental viewpoint,
will enable us to obtain analytical results. Insofar as the performances
of the medium are essentially determined by the contrast between its
maximum and minimum of transmission \cite{ma05}, it may be reasonably
expected that these results have some generality. Anticipating a more
extensive discussion, we note here that contrast exceeding 40 dB has
been actually used in absorbing media \cite{se85} whereas it seems
difficult to use an equivalent optical gain without generating serious
instabilities \cite{ste94,ste03b}.

The arrangement of our paper is as follows. In Sec.\ref{sec:GENERAL-ANALYSIS},
we introduce the basic equations of the problem and analyse their
physical meaning by privileging the temporal approach. Asymptotic
solutions obtained in the short and long pulse limits are given in
Sec.\ref{sec:SHORT-PULSE-LIMIT} and Sec.\ref{sec:LONG-PULSE}. Section
\ref{sec:OPTIMIZATION} is devoted to the optimization of the fractional
advance of the pulse maximum, ratio of this advance to the pulse duration.
The results are discussed in Sec.\ref{sec:DISCUSION} and we conclude
in Sec.\ref{sec:CONCLUSION} by summarizing the main points.

\section{GENERAL ANALYSIS\label{sec:GENERAL-ANALYSIS} }

When coherent light pulses with a slowly varying envelope propagate in
a linear medium of thickness $\ell$, the envelope $e(\ell,t)$ of
the output pulse can be deduced from the envelope $e(0,t)$ of the
incident pulse by the general relation:

\begin{equation}
e\left(\ell,t\right)=h\left(\ell,t\right)\otimes e\left(0,t\right)\label{eq:Un}
\end{equation}
where $h\left(\ell,t\right)$ is the impulse response of the medium
and $\otimes$ designates a convolution product \cite{pa87}. We consider
here a dilute medium with an isolated Lorentzian absorption-line at
a frequency $\omega_{0}$ coinciding with the carrier frequency $\omega_{c}$
of the incident pulses. We have then \cite{cri70,re1} 
\begin{equation}
h\left(\ell,t\right)=\delta\left(t\right)-L\gamma k\left(\ell,t\right)u_{H}\left(t\right)\label{eq:Deux}
\end{equation}
with 

\begin{equation}
k\left(\ell,t\right)=\frac{J_{1}\left(2\sqrt{L\gamma t}\right)}{\sqrt{L\gamma t}}e^{-\gamma t}.\label{eq:Trois}
\end{equation}
In these expressions, $\delta\left(t\right)$, $u_{H}\left(t\right)$
and $J_{1}\left(x\right)$ respectively designate the delta function,
the Heaviside unit step and the Bessel function of the first kind
of index 1; $t$ is the time retarded by the transit time $\ell/c$
at the velocity of light in vacuum (retarded time picture), $\gamma$
is the half-width at half-maximum of the absorption line and $L=\alpha\ell/2$
where $\alpha$ is the resonance absorption coefficient for the intensity.
From Eq.(\ref{eq:Un}), we finally get
\begin{equation}
e\left(\ell,t\right)=e\left(0,t\right)-L\gamma\left[k\left(\ell,t\right)u_{H}\left(t\right)\right]\otimes\left[e\left(0,t\right)\right].\label{eq:Quatre}
\end{equation}
From a physical viewpoint, Eq.(\ref{eq:Quatre}) first shows that,
if $e\left(0,t\right)$ starts at a given time, it is the same for
$e\left(\ell,t\right)$ at the corresponding retarded time in agreement
with relativistic causality. Second, as noted by Feynman in a more
general context \cite{fe63}, the transmitted wave is the sum of the
incident wave as if it had propagated in vacuum and of the wave reemitted
by the medium. Fast light thus results from the interference between
these two waves \cite{re2,bu04,ma08,cha09}.

In the frequency domain, the counterpart of Eq.(\ref{eq:Un}) reads
\begin{equation}
E\left(\ell,\Omega\right)=H\left(\ell,\Omega\right)E\left(0,\Omega\right)\label{eq:Cinq}
\end{equation}
where $E\left(\ell,\Omega\right)$, $H\left(\ell,\Omega\right)$ and
$E\left(0,\Omega\right)$ are, respectively, the Fourier transforms
of $e\left(\ell,t\right)$, $h\left(\ell,t\right)$ and $e\left(0,t\right)$.
$H\left(\ell,\Omega\right)$ is the transfer function of the medium
\cite{pa87} which in the present case has the simple form \cite{ma03}
\begin{equation}
H\left(\ell,\Omega\right)=\exp\left(-\frac{L}{1+i\Omega/\gamma}\right)\label{eq:Six}
\end{equation}
It yields a contrast $C$ between the maximum and the minimum of the
intensity transmission and a relative depth $D$ of the transmission
dip
\begin{equation}
C=\left|\frac{H\left(\ell,\infty\right)}{H\left(\ell,0\right)}\right|^{2}=e^{2L}=e^{\alpha\ell}\label{eq:sept}
\end{equation}
\begin{equation}
D=1-e^{-2L}=1-e^{-\alpha\ell}\label{eq:huit}
\end{equation}
We will restrict our analysis to contrasts below or equal to $50\,dB$
{[}$C_{dB}=10\log_{10}\left(C\right)${]} , which seems an upper limit
to the contrast that can be actually used in the experiments to avoid
parasitic effects. This contrast is attained for $L=5.756$ and the
corresponding dip depth $D$ is very close to $100\,\%$. Note that,
in most experimental demonstrations of fast light in atomic media,
this depth does not exceed the value $86\,\%$ attained in absorbing
media for the moderate optical thickness $\alpha\ell=2$ ($L=1$,
$C\approx8.7$ dB). 

General properties of the envelope of the transmitted field can be
deduced from the transfer function. We note first that $E\left(\ell,0\right)$
and $E\left(0,0\right)$ are, respectively, the areas of $e\left(\ell,t\right)$
and $e\left(0,t\right)$. The former is thus equal to the latter multiplied
by $e^{-L}$. Second, the mean time of $h\left(\ell,t\right)$, i.e.
the group delay $\tau_{g}$, equals the first cumulant $\kappa_{1}$
of $H\left(\ell,\Omega\right)$ \cite{do03}. Quite generally the
cumulants $\kappa_{n}$ are given by the following expansion of $\ln\left[H\left(\ell,\Omega\right)\right]$
: 
\begin{equation}
\ln\left[H\left(\ell,\Omega\right)\right]=\ln\left[H\left(\ell,0\right)\right]+\sum_{n=1}^{\infty}\frac{\kappa_{n}}{n!}\left(-i\Omega\right)^{n}.\label{eq:NeufA}
\end{equation}
In the present case, they take the simple form $\kappa_{n}=-n!L/\gamma^{n}$
and the group advance $a_{g}=-\tau_{g}$ reads : 
\begin{equation}
a_{g}=L/\gamma\label{eq:Neuf}
\end{equation}
Due the additivity property of the cumulants, $a_{g}$ is the advance
of the centre-of-gravity of $e\left(\ell,t\right)$ over that of $e\left(0,t\right)$,
that is, in our retarded time picture, over that of the envelope of
a pulse propagating without distortion at the velocity $c$. We emphasize
that this result is general and holds even when the transmitted pulse
is strongly distorted. As shows Eq.(\ref{eq:Neuf}), the advance $a_{g}$
is always positive. It should be carefully distinguished from the
advance $a$ of the pulse maximum and from that of the centre-of-gravity
of the intensity profile $\left|e\left(\ell,t\right)\right|^{2}$
which both may take negative values (delay instead of advance) \cite{pe00,ta01,ta05,ko05,na09}. 

\begin{figure}[h]
\centering{} \includegraphics[width=0.95\columnwidth]{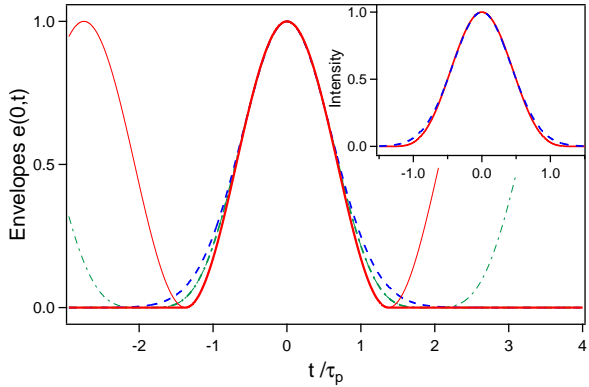}

\caption{Comparison of the envelopes $e_{1}(0,t)$ (solid red line), $e_{G}(0,t)$
(dashed blue line) and $e_{2}(0,t)$ (dash-dotted green line) of the
incident pulses for a same full width at half maximum $\tau_{p}$
of the corresponding intensity profiles. The thin lines are the periodic
continuations of $e_{1}(0,t)$ (solid red line) and $e_{2}(0,t)$
(dash-dotted green line). Inset: the corresponding intensity profiles
$\left|e_{1}(0,t)\right|^{2}$ (solid red line) and $\left|e_{G}(0,t)\right|^{2}$
(dashed blue line). The intensity profile $\left|e_{2}(0,t)\right|^{2}$
(not shown) is quasi confused with the Gaussian profile $\left|e_{G}(0,t)\right|^{2}$.\label{fig:Figure1}}
\end{figure}

In the following, we consider incident pulses of strictly finite duration,
the envelopes of which are good approximations of the Gaussian envelope
$e_{G}\left(0,t\right)=\exp\left(-t^{2}/2\sigma^{2}\right)$ usually
considered in the literature. We have retained
\begin{multline}
e_{1}\left(0,t\right)=\cos^{2}\left(\frac{\Omega_{1}t}{2}\right)\Pi\left(\frac{\Omega_{1}t}{2\pi}\right)\\
=\left[\frac{1+\cos\left(\Omega_{1}t\right)}{2}\right]\Pi\left(\frac{\Omega_{1}t}{2\pi}\right)\label{eq:Dix}
\end{multline}
or, if necessary, the more speculative form 
\begin{multline}
e_{2}\left(0,t\right)=\cos^{4}\left(\frac{\Omega_{2}t}{2}\right)\Pi\left(\frac{\Omega_{2}t}{2\pi}\right)\\
=\left[\frac{1+\cos\left(\Omega_{2}t\right)}{2}\right]^{2}\Pi\left(\frac{\Omega_{2}t}{2\pi}\right).\label{eq:Onze}
\end{multline}
In these expressions $\Pi\left(x\right)$ is the rectangle function
equal to 1 for $-1/2<x<1/2$ and 0 elsewhere. The envelopes $e_{1}\left(0,t\right)$,
$e_{G}\left(0,t\right)$ and $e_{2}\left(0,t\right)$ are compared
Fig.\ref{fig:Figure1} also as the corresponding intensity profiles
(nearly undistinguishable). 
The parameters $\Omega_{1}$, $\sigma$ and $\Omega_{2}$ are chosen
such that the corresponding intensity profiles (see inset in Fig.\ref{fig:Figure1})
have \emph{the same full width at half-maximum} $\tau_{p}$. This
is achieved when $\Omega_{1}\tau_{p}=4\,\arccos\left(2^{-1/4}\right)$,
$\tau_{p}=2\sigma\sqrt{\ln2}$ and $\Omega_{2}\tau_{p}=4\,\arccos\left(2^{-1/8}\right)$.\emph{
The duration} $\tau_{p}$\emph{ will be taken as universal time unit
in all the following}. In this way the profile of the transmitted
pulse is entirely determined by the two independent and dimensionless
parameters $\gamma\tau_{p}$ and $L$ (more generally the contrast
$C$). As above mentioned the challenge in fast light experiments
is to attain significant advance $a$ of the pulse maximum compared
to the pulse duration. From this viewpoint, a merit factor of the
experiments is the fractional advance $F=a/\tau_{p}$. Insofar as
the envelopes $e_{i}(0,t)$ ($i=1,2$ ) start at a finite time ($t=-\pi/\Omega_{i}$
), it is obvious that the fractional advance obtained with these envelopes
will reproduce that obtained with Gaussian pulses if and only if its
value is significantly less than $\pi/\left(\Omega_{i}\tau_{p}\right)$,
that is $1.37$ for $i=1$ and $1.91$ for $i=2$. As shown later
$e_{1}(0,t)$ constitute a sufficient approximation of $e_{G}(0,t)$
when $F\leq50\,\%$ whereas the better approximation provided by $e_{2}(0,t)$
is necessary for larger fractional advances. We finally mention that,
beyond the advantage of a finite duration, both envelopes offer the
possibility of a periodic continuation which will be useful to optimise
the fractional advance (Sec. \ref{sec:OPTIMIZATION}). Except when
contrary specified the envelope $e_{1}(0,t)$, leading to simpler
calculations, is used in the following.

\section{SHORT PULSE LIMIT\label{sec:SHORT-PULSE-LIMIT}}

When the incident pulse is short enough, its spectrum mainly lies
in the far wings of the line profile where the medium dispersion is
normal \cite{bo02}. A delay instead of an advance of the pulse maximum
might be expected in such conditions. We show in this section that\emph{
the opposite occurs}. Simple analytical results are obtained when
the duration $\tau_{p}$ of the incident pulse is short compared to
the duration of the pulse reemitted by the medium as considered in
the discussion of Eq.(\ref{eq:Quatre}) that is when $\gamma\tau_{p}\left(1+L/2\right)\ll1$
\cite{re3}. We have then
\begin{equation}
\left[k\left(\ell,t\right)u_{H}\left(t\right)\right]\otimes\left[e\left(0,t\right)\right]\approx k\left(\ell,t\right)\left[u_{H}\left(t\right)\otimes e\left(0,t\right)\right].\label{eq:Douze}
\end{equation}
For an incident pulse of envelope $e_{1}(0,t)$, it is easily shown
that
\begin{multline}
\left[u_{H}\left(t\right)\otimes e\left(0,t\right)\right]=\frac{1}{2}\left[t+\frac{\sin\left(\Omega_{1}t\right)}{\Omega_{1}}+\frac{\pi}{\Omega_{1}}\right]\Pi\left(\frac{\Omega_{1}t}{2\pi}\right)\\
+\frac{\pi}{\Omega_{1}}u_{H}\left(t-\frac{\pi}{\Omega_{1}}\right)\label{eq:Treize}
\end{multline}
and, according to Eq.(\ref{eq:Quatre}), the envelope of the transmitted
field reads
\begin{equation}
e_{1}(\ell,t)\approx e_{1}(0,t)-L\gamma\left[u_{H}\left(t\right)\otimes e\left(0,t\right)\right]\frac{J_{1}\left(2\sqrt{L\gamma t}\right)}{\sqrt{L\gamma t}}e^{-\gamma t}.\label{eq:Quatorze}
\end{equation}

\begin{figure}[h]
\centering{} \includegraphics[width=0.95\columnwidth]{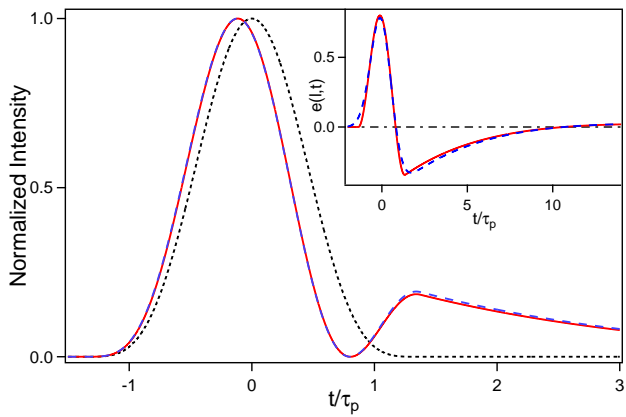} \caption{Normalized intensity profiles of the transmitted pulse obtained for
$e(0,t)=e_{1}(0,t)$ , $L=5.756$ (transmission contrast $C=50$ dB
) and $\tau_{p}=0.06/\gamma$ ($\Omega_{1}=38\gamma$ ). The solid
red line (the dashed blue line) is the exact solution obtained by
FFT (the analytical solution obtained in the short pulse limit). As
in all the following figures, the intensity profile of the incident
pulse (dotted black line) is given for reference. Its full width at
half maximum $\tau_{p}$ is taken as time unit. Inset: comparison
of the envelopes of the transmitted pulse obtained for $e(0,t)=e_{1}(0,t)$
(solid red line) and $e(0,t)=e_{G}(0,t)$ (dashed blue line). The
pulse tail displays a first zero for $t/\tau_{p}\approx j_{1,1}^{2}/\left(4L\gamma\tau_{p}\right)\approx10.6$
(see text). \label{fig:Figure2}}
\end{figure}

\begin{figure}
\centering{} \includegraphics[width=0.95\columnwidth]{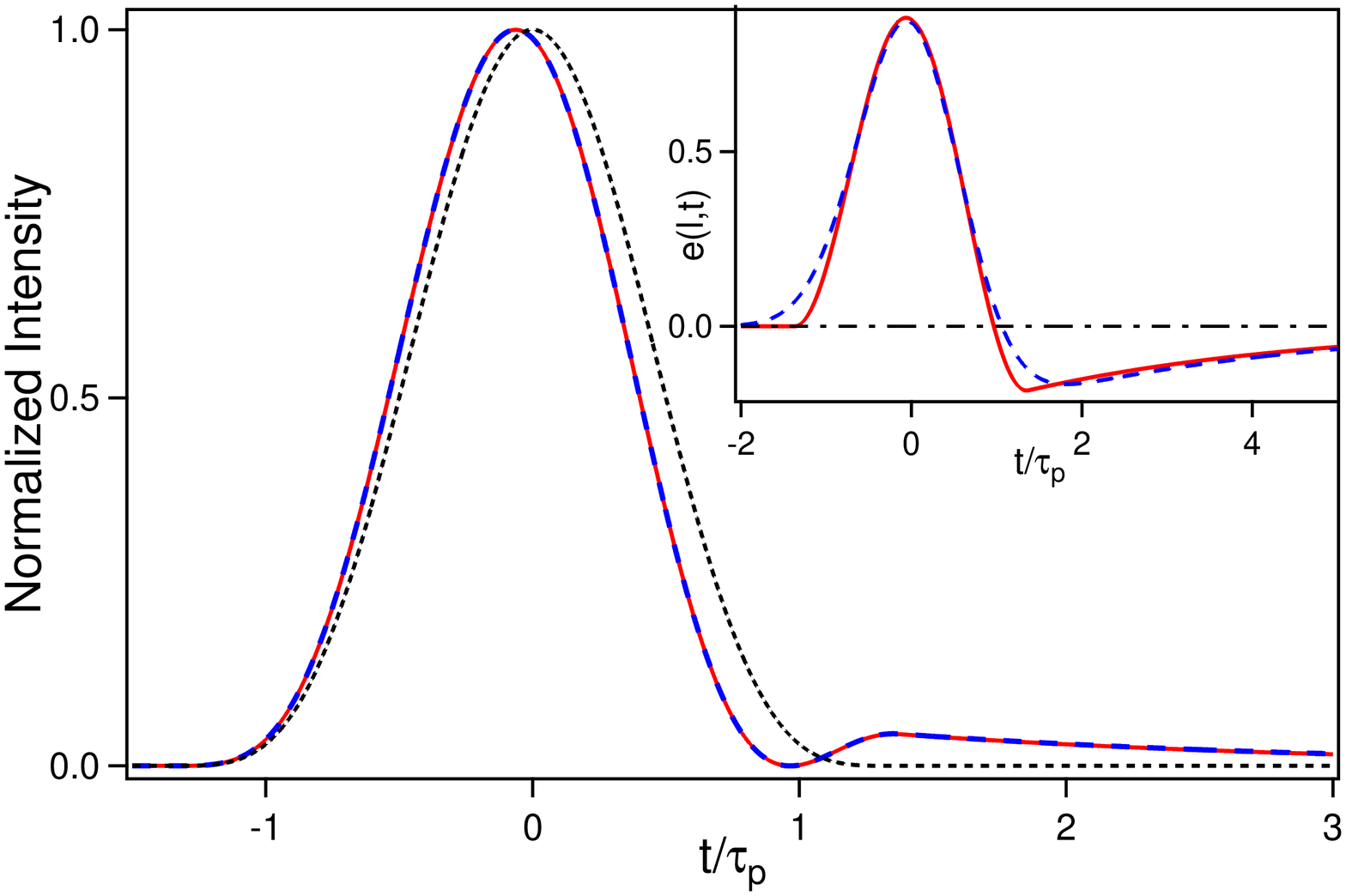} \caption{Same as Fig \ref{fig:Figure2} when $L=1$ ($C\approx8.7$ dB, dip
depth $D\approx86\%$) and $\tau_{p}=0.2/\gamma$ ($\Omega_{1}=11.4\gamma$
). The tail of the envelopes (inset) has now an exponential decrease
(see text).\label{fig:Figure3}}
\end{figure}

Figures \ref{fig:Figure2} and \ref{fig:Figure3} are respectively
obtained for large ($L=5.756$) and moderate ($L=1$) optical thickness.
They show that the intensity profiles $\left|e_{1}\left(\ell,t\right)\right|^{2}$
determined by Eq.(\ref{eq:Quatorze}) are in perfect agreement with
the exact profiles derived by fast Fourier transform (FFT). 
Quite generally the fractional advance $F=a/\tau_{p}$ of the pulse
maximum has the approximate form
\begin{equation}
F\approx\frac{2L\gamma}{\Omega_{1}^{2}\tau_{p}}=\frac{2L\gamma\tau_{p}}{\Omega_{1}^{2}\tau_{p}^{2}}\approx0.38L\gamma\tau_{p}\label{eq:Quinze}
\end{equation}
whereas $a/\tau_{p}=L/\gamma\tau_{p}$ and thus $a/a_{g}=0.38\left(\gamma\tau_{p}\right)^{2}$.
This means that, in the short pulse limit, the advance of the pulse
maximum will be always much smaller than the group advance. This apparent
paradox is explained by the fact that the envelope $e_{1}(\ell,t)$
has a long tail which is essentially negative and artificially increases
the advance $a_{g}$ of its centre of gravity. This tail has the general
form 
\begin{equation}
e_{1}(\ell,t)\approx-\frac{\pi L\gamma}{\Omega_{1}}k\left(\ell,t\right)\approx-1.37L\gamma\tau_{p}k\left(\ell,t\right)\label{eq:Seize}
\end{equation}
For large optical thickness, it oscillates with successive lobes of
rapidly decreasing amplitude and zeroes at the times $t_{n}=j_{1,n}^{2}/\left(4L\gamma\right)$
where $j_{1,n}$ designates the corresponding zeroes of $J_{1}\left(x\right)$.
This behaviour is illustrated in the inset of Fig.\ref{fig:Figure2}.
Note that, for the parameters considered, the amplitude of the secondary
lobes appearing for $t>t_{1}$ (not shown) is very weak, below $0.02$
for the first and largest one. For moderate optical thickness, the
decrease of $k\left(\ell,t\right)$ is mainly determined by the term
$\exp\left(-\gamma t\right)$ and the tail of the envelope is reduced
to a simple exponential of the form $-1.37L\gamma\tau_{p}\exp\left[-\gamma\left(1+L/2\right)t\right]$
as shown in the inset of Fig.\ref{fig:Figure3} obtained for $L=1$.
For a five times smaller optical thickness ($L=0.2$), the intensity
of the tail would be 25 times weaker and the fractional advance of
the maximum 5 times smaller. The normalized intensity profile of the
transmitted pulse is then nearly undistinguishable from that of the
incident pulse although the depth of the dip in the transmission profile
is significant ($D=33\,\%$).

Similar calculations can be performed for Gaussian pulses. We only
give below the expression of the envelope $e_{G}(\ell,t)$ of the
transmitted pulse:
\begin{equation}
e_{G}(\ell,t)\approx\exp\left(-\frac{t^{2}}{2\sigma^{2}}\right)-L\gamma\sigma\sqrt{\frac{\pi}{2}}\left[1+\mathrm{erf}\left(\frac{t}{\sigma\sqrt{2}}\right)\right]k\left(\ell,t\right)\label{eq:Dixsept}
\end{equation}
where $\mathrm{erf}\left(x\right)$ designates the error function.
There is again a perfect agreement with the exact numerical result
obtained by FFT. The fractional advance has now the approximate form
\begin{equation}
F\approx\frac{L\gamma\sigma^{2}}{\tau_{p}}=\left(\frac{\sigma}{\tau_{p}}\right)^{2}L\gamma\tau_{p}\approx0.36L\gamma\tau_{p}\label{eq:DixHuit}
\end{equation}
 The envelopes $e_{G}(\ell,t)$ and $e_{1}(\ell,t)$ are shown in
the insets of Fig.\ref{fig:Figure2} and Fig.\ref{fig:Figure3}. As
expected, they are quite comparable. 

\section{LONG PULSE OR ADIABATIC LIMIT\label{sec:LONG-PULSE}}

We consider in this section the case opposite to the previous one.
We assume that the envelope $e(0,t)$ of the incident pulse is \emph{everywhere}
slowly varying compared to $k\left(\ell,t\right)u_{H}\left(t\right)$.
This implies in particular that the pulse duration $\tau_{p}$ is
long compared to $1/\gamma$. The function $L\gamma k\left(\ell,t\right)u_{H}\left(t\right)$
can then be approximated by $A_{k}\delta\left(t-\tau_{kg}\right)$
where $A_{k}$ and $\tau_{kg}$ are, respectively, its area and its
centre of gravity. Since $L\gamma k\left(\ell,t\right)u_{H}\left(t\right)$
is the inverse Fourier transform of $K\left(\ell,\Omega\right)=1-H\left(\ell,\Omega\right)$,
we have $A_{k}=1-H\left(\ell,0\right)=1-e^{-L}$ and, by an expansion
of $\ln\left[K\left(\ell,\Omega\right)\right]$ in cumulants \cite{do03},
$\tau_{kg}=L/\left[\gamma\left(e^{L}-1\right)\right]>0$. Injecting
these results in Eq.(\ref{eq:Quatre}), we get 
\begin{equation}
e\left(\ell,t\right)\approx e\left(0,t\right)-A_{k}e\left(0,t-\tau_{kg}\right)\label{eq:Dixneuf}
\end{equation}
 The envelope of the output field appears equal to that of the input
field minus its copy reduced in amplitude and slightly retarded \cite{re2,ma08,cha09}.
Assuming that $\tau_{kg}\ll\tau_{p}$, Eq.(\ref{eq:Dixneuf}) takes
the form
\begin{equation}
e\left(\ell,t\right)\approx e^{-L}\left[e\left(0,t\right)+a_{g}\dot{e}\left(0,t\right)\right]\label{eq:vingt}
\end{equation}
where the dot designates a time-derivative and where we have taken
into account the relations $1-A_{k}=e^{-L}$ and $A_{k}\tau_{kg}/\left(1-A_{k}\right)=a_{g}$.
In the limit where $a_{g}/\tau_{p}\ll1$, we retrieve the classical
result 
\begin{equation}
e\left(\ell,t\right)\approx e^{-L}e\left(0,t+a_{g}\right)\label{eq:vingtetun}
\end{equation}
In this limit, the envelope of the output pulse equals that of the
input pulse attenuated by the steady state transmission $e^{-L}$
for the field and advanced by $a_{g}$ (no distortion). It should
be however kept in mind that this asymptotic result only holds when
$a_{g}/\tau_{p}\ll1$ and, consequently, that the corresponding fractional
delay will be limited. 
\begin{figure}
\centering{} \includegraphics[width=0.95\columnwidth]{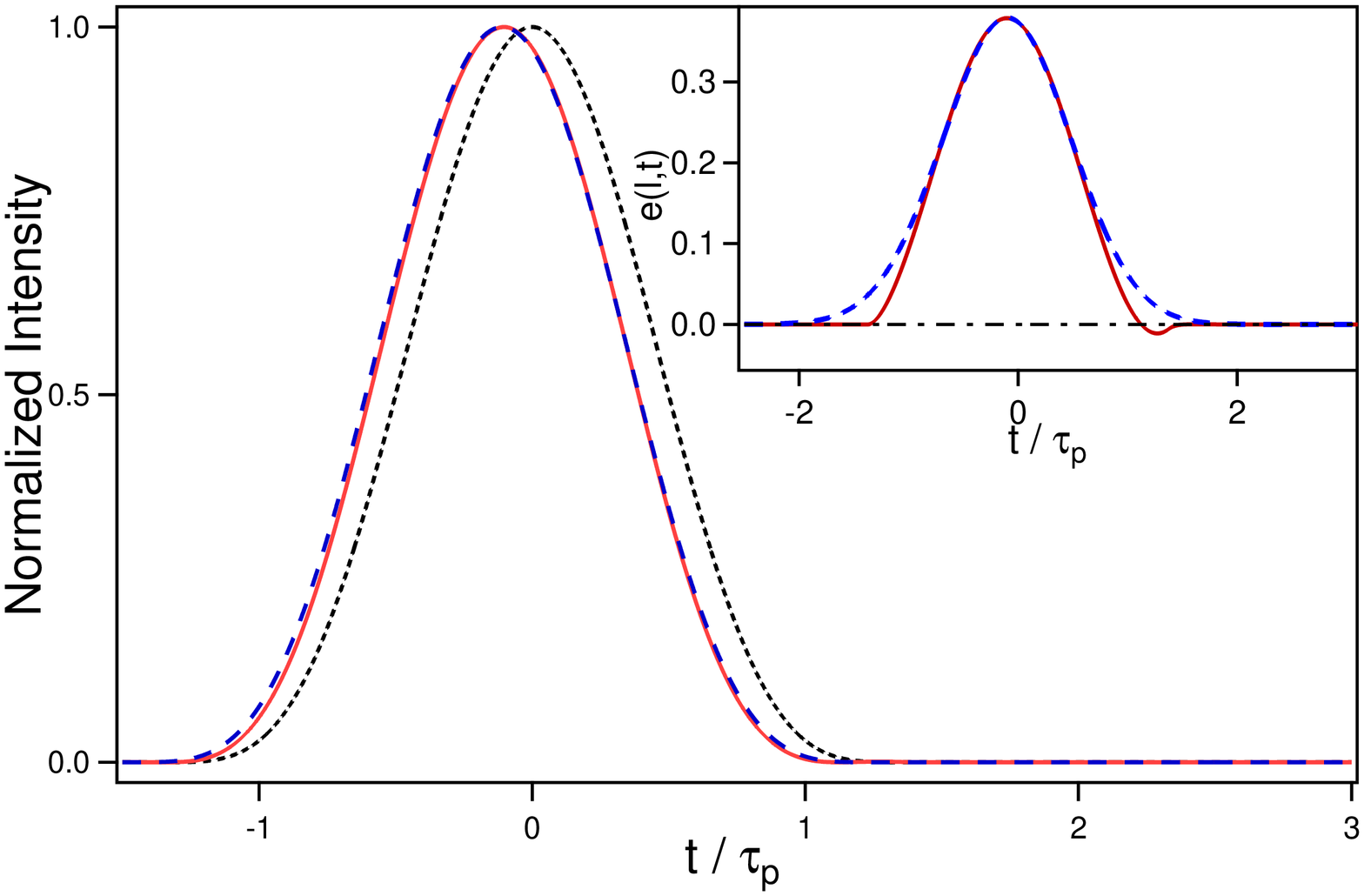} \caption{Normalized intensity profiles of the transmitted pulse obtained for
$e(0,t)=e_{1}(0,t)$ , $L=1$ ($C=8.7$ dB, dip depth $D\approx86\%$
) and $\Omega_{1}=\gamma/4$ ($\tau_{p}\approx9.15/\gamma$). The
solid red line (the dashed blue line) is the exact solution obtained
by FFT (the analytical solution obtained in the long pulse or adiabatic
limit). Inset: comparison of the envelopes of the transmitted pulse
obtained for $e(0,t)=e_{1}(0,t)$ (solid red line) and $e(0,t)=e_{G}(0,t)$
(dashed blue line).\label{fig:Figure4}}
\end{figure}

Figure \ref{fig:Figure4} shows an example of situation for which
Eq.(\ref{eq:vingtetun}) provides a good approximation of the exact
result. It is obtained for $e(0,t)=e_{1}(0,t)$ and the moderate optical
thickness already considered in Fig.\ref{fig:Figure3} ($L=1$ leading
to $C\approx8.7\,dB$ and $D\approx86\,\%$) but a much longer pulse
duration ($\Omega_{1}/\gamma=1/4$ that is $\gamma\tau_{p}\approx9.1$).
The fractional advance $F=a/\tau_{p}$ of the pulse maximum is equal
to $10.3\,\%$ whereas $a/a_{g}=0.94$, actually close to unity. 

The asymptotic solution given by Eq.(\ref{eq:vingtetun}) remains
acceptable for long pulses even when the condition $a_{g}\ll\tau_{p}$
is poorly satisfied.
\begin{figure}
\centering{} \includegraphics[width=0.95\columnwidth]{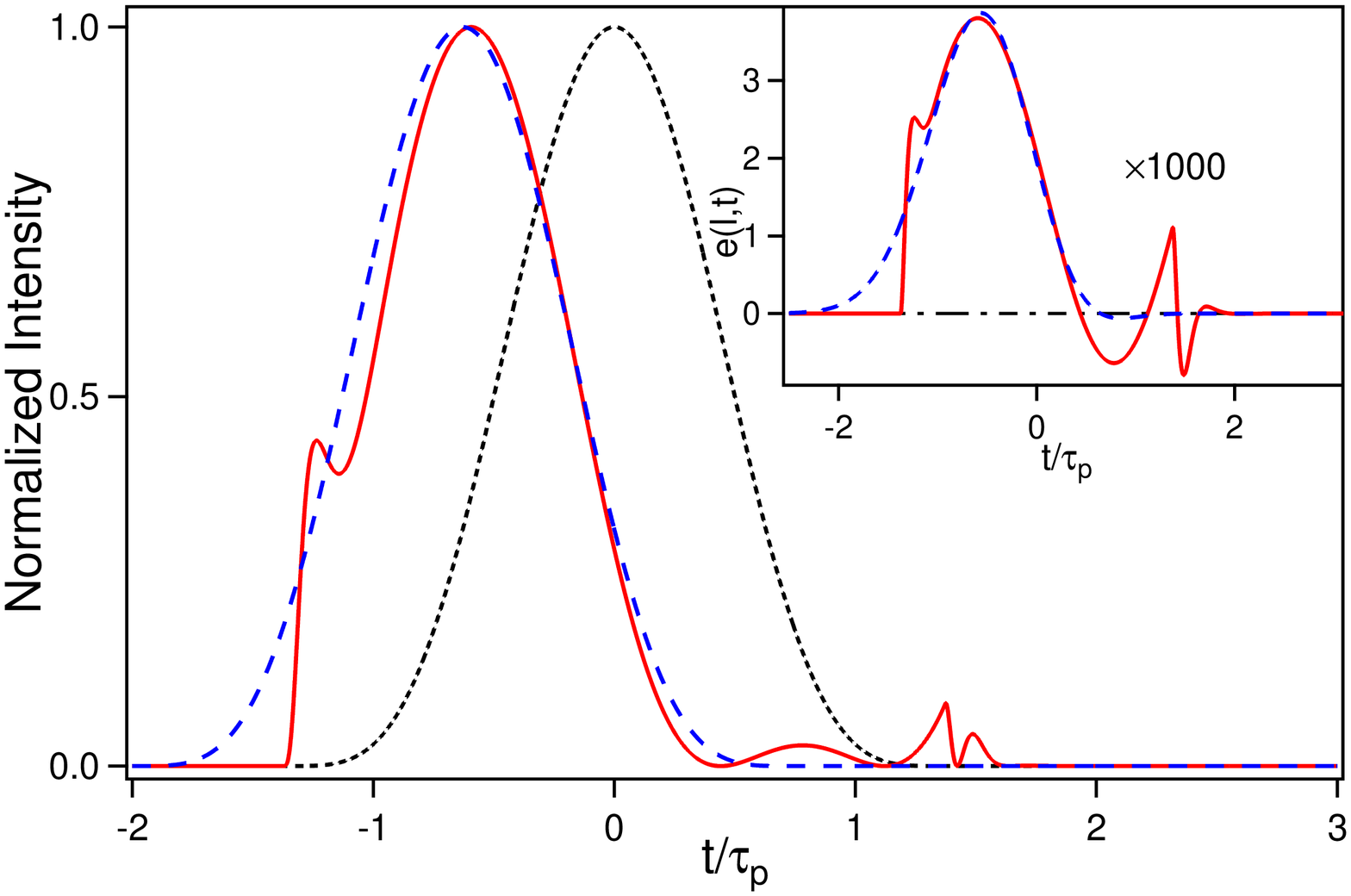} \caption{Same as Fig \ref{fig:Figure4} when $L=5.756$ ($C=50$ dB ) instead
of $L=1$. Note the presence of a precursor and a postcursor which
had negligible amplitudes for $L=1$ and are obviously absent in the
case of a Gaussian incident pulse (see inset).\label{fig:Figure5}}
\end{figure}
Figure \ref{fig:Figure5} illustrates this point. It is also obtained
for $\gamma\tau_{p}\approx9.1$ but for $L=5.756$ ($C=50\,dB$).
We have now a fractional advance $F=a/\tau_{p}\approx59\,\%$ and,
despite that the condition of validity of Eq.(\ref{eq:vingtetun})
is far from being fulfilled ($a_{g}/\tau_{p}\approx0.63$), $a$ and
$a_{g}$ remain close. A new feature of Fig.\ref{fig:Figure5} compared
to Fig.\ref{fig:Figure4} is the appearance of transients at the beginning
(precursor) and the end (postcursor) of the envelope $e_{1}(0,t)$,
that is for $t/\tau_{p}=\pm\pi/\Omega_{1}\tau_{p}\approx\pm1.37$.
They obviously originate in the discontinuity of the second derivative
of $e_{1}(0,t)$ at the corresponding times, are absent in the case
of Gaussian pulses (see inset of Fig.\ref{fig:Figure5}) and are not
taken into account by the adiabatic approximation that assumes that
the envelope of the incident pulse is \emph{everywhere} continuous
with continuous derivatives. By a cavalier extrapolation of results
obtained in the study of the Sommerfeld precursors \cite{ma12}, we
have found that the peak amplitude of the precursor roughly scales
as 
\begin{equation}
A_{p}=\left(\frac{\Omega_{1}}{\gamma L}\right)^{2}\label{eq:vingtdeux}
\end{equation}
We have numerically checked that this law provides a good estimate
of the exact result as soon as the precursor is visible without dominating
excessively the main pulse. In the conditions of Fig.\ref{fig:Figure5},
we get $A_{p}=1.9\times10^{-3}$ in satisfactory agreement with the
amplitude ($\approx2.5\times10^{-3}$) of the precursor shown in the
inset of this figure\textbf{. }The order of magnitude given by Eq.(\ref{eq:vingtdeux})
also applies to the postcursor but there is now interference of the
postcursor with the tail of the main field and it is then difficult
to determine what originates in the postcursor alone.

\section{OPTIMIZATION OF THE FRACTIONAL PULSE-ADVANCE\label{sec:OPTIMIZATION}}

The convolution approach used in the previous sections provides both
a physical insight on the fast light phenomenon and analytical results
in the short and long pulse limits. Unfortunately the corresponding
fractional advances $F=a/\tau_{p}$ may be weak. Larger fractional
advances are expected for intermediate pulse durations. The convolution
approach can be extended to this case when the optical thickness is
moderate, say for $L\leq1$. The function $k\left(\ell,t\right)$
appearing in the impulse response may then be approximated by $\exp\left[-\gamma t\left(1+L/2\right)\right]$
and the convolution product can be explicitly calculated. We will
not develop this method here. Indeed more general and simpler results
on $e(\ell,t)$, can be derived from the envelope $\widehat{e}(\ell,t)$
of the transmitted field obtained when the envelope $e(0,t)$ of the
incident pulse is replaced by its periodic continuation $\widehat{e}(0,t)$
(see the thin lines in Fig.\ref{fig:Figure1}).

When $e(0,t)=e_{1}(0,t)$, its periodic continuation $\widehat{e}_{1}(0,t)=\left[1+\cos\left(\Omega_{1}t\right)\right]/2$
contains only three frequencies ($0$, $\pm\Omega_{1}$). Taking into
account that $H\left(\ell,\Omega\right)$ and $H\left(\ell,-\Omega\right)$
are complex conjugates, we get 
\begin{equation}
\widehat{e}_{1}(\ell,t)=\frac{\exp\left(-L\right)}{2}+\frac{\left|H\left(\ell,\Omega_{1}\right)\right|}{2}\cos\left[\Omega_{1}t+\Phi\left(\ell,\Omega_{1}\right)\right]\label{eq:vingttrois}
\end{equation}
where 
\begin{equation}
\Phi\left(\ell,\Omega\right)=\arg\left[H\left(\ell,\Omega\right)\right]=\frac{L\Omega/\gamma}{1+\Omega^{2}/\gamma^{2}}.\label{eq:vingtquatre}
\end{equation}
The advance $\widehat{a}_{1}$ of the maximum of $\widehat{e}_{1}(\ell,t)$
on that of $\widehat{e}_{1}(0,t)$ is thus
\begin{equation}
\widehat{a}_{1}=\frac{\Phi\left(\ell,\Omega_{1}\right)}{\Omega_{1}}=\frac{L/\gamma}{1+\Omega_{1}^{2}/\gamma^{2}}=\frac{a_{g}}{1+\Omega_{1}^{2}/\gamma^{2}}.\label{eq:vingtcinq}
\end{equation}
So long as the responses to the successive pulses of $\widehat{e}_{1}(0,t)$
do not significantly overlap and as the precursor and postcursor have
a negligible amplitude, $\widehat{e}_{1}(\ell,t)$ is expected to
be a good approximation of $e_{1}(\ell,t)$ in their common domain
of existence. We have then $a\approx\widehat{a}_{1}$ and the fractional
advance reads
\begin{equation}
F=\frac{a}{\tau_{p}}\approx\frac{\Omega_{1}\widehat{a}_{1}}{\Omega_{1}\tau_{p}}\approx0.44\Omega_{1}\widehat{a}_{1}=0.44\frac{L\Omega_{1}/\gamma}{1+\Omega_{1}^{2}/\gamma^{2}}.\label{eq:vingtsix}
\end{equation}
Equation (\ref{eq:vingtcinq}) shows that the advance is the largest
and equal to the group advance $a_{g}$ when $\Omega_{1}^{2}/\gamma^{2}\ll1$,
that is for long incident pulses. Unfortunately, it results from Eq.(\ref{eq:vingtsix})
that the corresponding fractional advance $F$ is very weak. $F$
attains its maximum value $F_{m}\approx0.22L$ when $\Omega_{1}=\gamma$
($\gamma\tau_{p}\approx2.29$). The corresponding advance is half
of the group advance ($a\approx a_{g}/2$). Besides the main pulse
is followed by a second one, the relative intensity of which reads
\begin{equation}
R=\tanh^{2}\left(L/4\right).\label{eq:vingtsept}
\end{equation}
\begin{figure}
\centering{}\emph{ }\includegraphics[width=0.95\columnwidth]{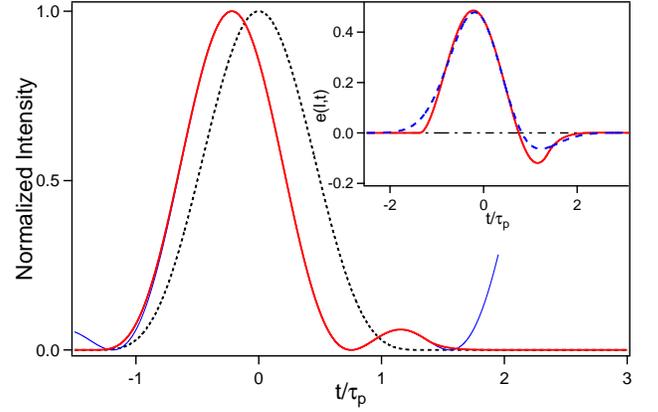}
\caption{Comparison for $L=1$ ($C=8.7$ dB, dip depth $D\approx86\%$) of
the exact intensity profile of the transmitted pulse (solid red line)
with the analytical profile obtained by periodically continuing the
incident pulse $e(0,t)=e_{1}(0,t)$ (thin blue line). The pulse duration
is that maximizing the fractional advance ($\Omega_{1}=\gamma$, $\tau_{p}\approx2.29/\gamma$)
. It leads to $a=a_{g}/2$ and to $F=L/2\gamma\tau_{p}\approx22\%$
. Inset: comparison of the envelopes of the transmitted pulse obtained
for $e(0,t)=e_{1}(0,t)$ (solid red line) and $e(0,t)=e_{G}(0,t)$
(dashed blue line).\label{fig:Figure6}}
\end{figure}
Figure \ref{fig:Figure6} obtained for $L=1$ and $\Omega_{1}=\gamma$,
shows that the exact intensity profile $\left|e_{1}(l,t)\right|^{2}$
is actually quasi-confused with its periodic approximation $\left|\widehat{e}_{1}(l,t)\right|^{2}$
. $F$ and $R$ are in perfect agreement with those given by Eq.(\ref{eq:vingtsix})
and Eq.(\ref{eq:vingtsept}), namely $F=F_{m}=22\,\%$ and $R=6\,\%$.
A simulation (not shown) made for $\Omega_{1}=\gamma$ and $L=2$
($C\approx17\,dB$) evidences that $\left|\widehat{e}_{1}(l,t)\right|^{2}$
is still a very good approximation of $\left|e_{1}(l,t)\right|^{2}$.
We get then $F\approx42\,\%$ and $R=22\,\%$, very close to the values
$F\approx44\,\%$ and $R=21\,\%$ predicted by Eq.(\ref{eq:vingtsix})
and Eq.(\ref{eq:vingtsept}). Note that the relative intensity of
the second maximum of intensity is significantly larger than that
obtained for $L=1$. 

For large optical thickness, problems arise from the overlapping of
the responses to the successive pulses of the periodically continued
input envelope and from the precursor and postcursor that can have
significant amplitudes. Both problems are solved by replacing $e_{1}(0,t)$
considered above by $e_{2}(0,t)$ which is a better approximation
of the Gaussian pulse. Indeed the successive pulses of $\widehat{e}_{2}(0,t)$
are better separated than those of $\widehat{e}_{1}(0,t)$ (see Fig.\ref{fig:Figure1})
and the higher order initial and final discontinuities of $e_{2}(0,t)$
ensure negligible amplitude of the precursor and postcursor. $\widehat{e}(0,t)$
reads 
\begin{equation}
\widehat{e}_{2}(0,t)=\cos^{4}\left(\frac{\Omega_{2}t}{2}\right)=\frac{3}{8}+\frac{\cos\left(\Omega_{2}t\right)}{2}+\frac{\cos\left(2\Omega_{2}t\right)}{8}.\label{eq:vingthuit}
\end{equation}
It contains five frequencies ($0$, $\pm\Omega_{2}$ and $\pm2\Omega_{2}$)
and leads to
\begin{multline}
\widehat{e}_{2}(\ell,t)=\frac{3}{8}\exp\left(-L\right)+\frac{\left|H\left(\ell,\Omega_{2}\right)\right|}{2}\cos\left[\Omega_{2}t+\Phi\left(\ell,\Omega_{2}\right)\right]\\
+\frac{\left|H\left(\ell,2\Omega_{2}\right)\right|}{8}\cos\left[2\Omega_{2}t+\Phi\left(\ell,2\Omega_{2}\right)\right].\label{eq:vingtneuf}
\end{multline}
The advance $\widehat{a}_{2}$ of the maximum of $\widehat{e}_{2}(\ell,t)$
is obtained by solving the transcendent equation derived from Eq.(\ref{eq:vingtneuf})
\begin{equation}
\frac{\sin\left[\Phi\left(\ell,2\Omega_{2}\right)-2\Omega_{2}\widehat{a}_{2}\right]}{\sin\left[\Phi\left(\ell,\Omega_{2}\right)-\Omega_{2}\widehat{a}_{2}\right]}=-2\left|\frac{H\left(\ell,\Omega_{2}\right)}{H\left(\ell,2\Omega_{2}\right)}\right|.\label{eq:trente}
\end{equation}
When $L=3$ ($C\approx26\,dB$), we find that $\widehat{F}=\widehat{a}_{2}/\tau_{p}$
takes its largest value $\widehat{F}_{m}\approx0.55$ for $\Omega_{2}\approx0.644\gamma$
($\gamma\tau_{p}\approx2.55$).
\begin{figure}
\centering\includegraphics[width=0.95\columnwidth]{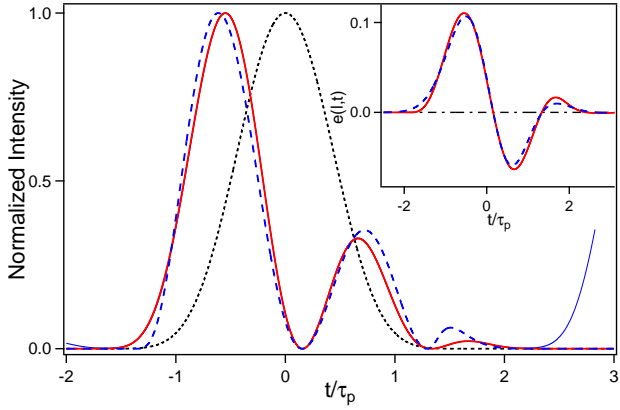} \caption{Comparison for $L=3$ ($C=26$ dB) of the exact intensity profiles
$\left|e_{2}(l,t)\right|^{2}$ (solid red line) and $\left|e_{1}(l,t)\right|^{2}$
(dashed blue line) with the analytical periodic profile $\left|\widehat{e}_{2}(l,t)\right|^{2}$
(thin blue line). The pulse duration ($\tau_{p}\approx2.55/\gamma$)
is that maximizing the fractional advance of $\left|\widehat{e}_{2}(l,t)\right|^{2}$.
Inset: comparison of the envelopes of the transmitted pulse obtained
for $e(0,t)=e_{2}(0,t)$ (solid red line) and for $e(0,t)=e_{G}(0,t)$
(dashed blue line).\label{fig:Figure7}}
\end{figure}
Figure \ref{fig:Figure7}, obtained in those conditions, shows that
the profile $\left|\widehat{e}_{2}(\ell,t)\right|^{2}$ perfectly
fits the exact intensity profile $\left|e_{2}(\ell,t)\right|^{2}$
with $F\approx\widehat{F}_{m}\approx55\,\%$, $a/a_{g}\approx0.45$
and a relative intensity of the second maximum $R\approx33\,\%$.
We have added on Fig.\ref{fig:Figure7} the profile $\left|e_{1}(\ell,t)\right|^{2}$
which would be obtained for the same duration of the incident pulse.
For this optical thickness, it does not deviate too strongly from
the previous one with $F\approx61\,\%$, $a/a_{g}\approx0.52$ and
$R\approx35\,\%$. The situation dramatically changes for the largest
transmission-contrast and optical thickness considered in this article
($C=50\,dB$, $L=5.756$). For $e(0,t)=e_{1}(0,t)$, the transmitted
pulse is then strongly damaged by the precursor and postcursor and
the periodic solution $\widehat{e}_{1}(l,t)$ fails to reproduce the
exact result (even very approximately). On the other hand, the periodic
solution given by Eq.(\ref{eq:vingtneuf}) remains a very good approximation
of the exact result when $e(0,t)=e_{2}(0,t)$. 
\begin{figure}
\centering{} \includegraphics[width=0.95\columnwidth]{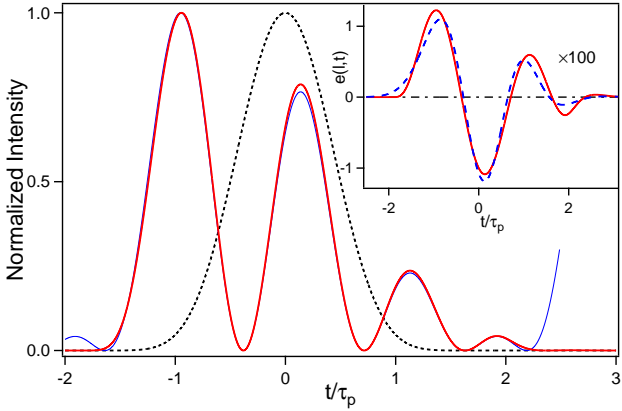}\caption{Comparison for $L=5.756$ ($C=50$ dB ) of the exact intensity profiles
s $\left|e_{2}(l,t)\right|^{2}$ (solid red line) with the analytical
periodic profile $\left|\widehat{e}_{2}(l,t)\right|^{2}$ (thin blue
line). The pulse duration is that maximizing the fractional advance,
that is $\tau_{p}\approx3.18/\gamma$ leading to $F\approx94\%$ and
$a=0.52\,a_{g}$. Inset: comparison of the envelopes of the transmitted
pulse obtained for $e(0,t)=e_{2}(0,t)$ (solid red line) and for $e(0,t)=e_{G}(0,t)$
(dashed blue line). Note that the slow light regime is attained for
the Gaussian pulse (see text).\label{fig:Figure8}}
\end{figure}
Figure \ref{fig:Figure8} shows the intensity profiles obtained for
the pulse duration $\gamma\tau_{p}\approx3.18$ ($\Omega_{2}\approx0.517\gamma$)
leading to the maximum fractional advance $F\approx\widehat{F}_{m}\approx94\,\%$
with an absolute advance $a\approx0.52a_{g}$. The only difference
between the exact intensity profile $\left|e_{2}(\ell,t)\right|^{2}$
and its analytical periodic counterpart $\left|\widehat{e}_{2}(\ell,t)\right|^{2}$
lies in the second lobe, the relative intensity ($R\approx79\,\%$)
of which is very slightly underestimated in $\left|\widehat{e}_{2}(\ell,t)\right|^{2}$. 

A main result of the above study is that the pulse following the main
one becomes invading when the transmission contrast $C$ increases.
For $C\geq25\,dB$ ($L\geq2.9$), there is a range of pulse durations
for which the peak intensity of the second pulse exceeds that of the
first one. There is then a transition from fast to slow light \cite{ta01,ta05,ko05,na09}.
When $C=50\,dB$ and $e(0,t)=e_{2}(0,t)$, this occurs for $0.185<\gamma\tau_{p}<2.95$.
For Gaussian incident pulses, the upper limit increases up to $\gamma\tau_{p}\approx3.34$
and the inset of Fig.\ref{fig:Figure8} (obtained for $\gamma\tau_{p}\approx3.18$)
shows that the slow light regime is actually attained. We however
remark that fast light is always obtained when the pulse duration
is sufficiently short or long.

As mentioned in our introduction the main challenge of fast light
experiments is to obtain a significant fractional advance with \emph{moderate
pulse distortion}. The first objective is attained in the conditions
of Fig.\ref{fig:Figure8} but not the latter owing to the presence
of large secondary lobes in the intensity profile of the transmitted
pulse. In order to reduce their relative intensity to a few percents,
the incident pulse should be lengthened and this obviously reduces
the fractional advance.
\begin{figure}
\centering{} \includegraphics[width=0.95\columnwidth]{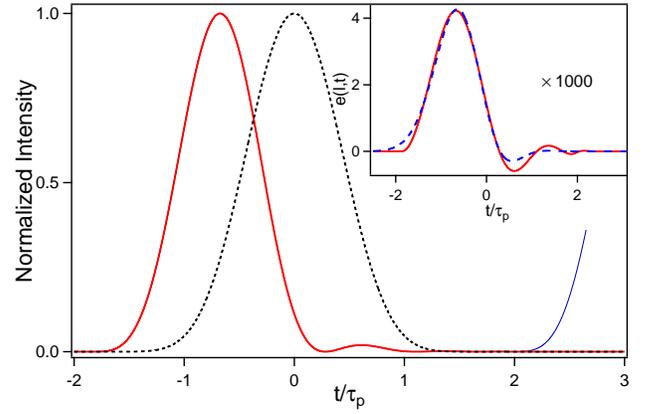} \caption{Same as Fig \ref{fig:Figure8} for $\tau_{p}\approx7.5/\gamma$ instead
of $3.18/\gamma$ . The relative intensity of the \emph{s}econd maximum
is so reduced to $R=2\%$ whereas the fractional advance falls to
$F\approx67\%$. The inset shows that $e_{2}(0,t)$ and $e_{G}(0,t)$
both lead to fast light with very close pulse advances.\label{fig:Figure9}}
\end{figure}
Figure \ref{fig:Figure9} is obtained in the conditions of Fig.\ref{fig:Figure8}
for $\gamma\tau_{p}=7.5$ instead of $3.18$. The relative intensity
of the second maximum (the fractional advance) then falls to $R\approx2\,\%$
($F\approx67\,\%$) whereas $a\approx0.88a_{g}$. Note that the intensity
profile of the transmitted pulse is perfectly fitted by the periodic
solution $\left|\widehat{e}_{2}(\ell,t)\right|^{2}$. This result
is general. As long as the periodic solution works for the pulse duration
maximizing the fractional delay, it works better for longer pulses.
In particular, the approximation $e\left(\ell,t\right)\approx e_{1}\left(\ell,t\right)$,
very good for $L=1$ and $\Omega_{1}=\gamma$ (Fig.\ref{fig:Figure6}),
is excellent when $L=1$ and $\Omega_{1}=\gamma/4$ (Fig.\ref{fig:Figure4}),
leading in this latter case to an advance which is exact with four
significant figures.

When the intensity of the secondary lobes is weak enough, the distortion
of the transmitted pulse mainly consists in a narrowing. This simply
results from the negative value of the second cumulant of $H\left(\ell,\Omega\right)$
and the additivity of the cumulants \cite{do03}. The narrowing of
the pulse originates a difference between the advances $a_{\uparrow}$
at its rise and $a_{\downarrow}$ at its fall. The distortion can
then conveniently characterized by the narrowing indicator $\xi=\left(a_{\downarrow}-a_{\uparrow}\right)/\left(a_{\downarrow}+a_{\uparrow}\right)$
where the advances $a_{\uparrow}$ and $a_{\downarrow}$ are measured
at half-maximum intensity \cite{ma03}. In the conditions of Fig.\ref{fig:Figure4}
{[}Fig.\ref{fig:Figure9}{]}, we get $\xi\approx8\,\%$ {[}$\xi\approx13\,\%${]}
in agreement with the value derived from Eq.(\ref{eq:vingttrois})
{[}Eq.(\ref{eq:vingtneuf}){]}. 

The procedure of periodic continuation used in this section to maximize
the fractional pulse advance can also be used to determine the effect
of a detuning of the carrier frequency $\omega_{c}$ of the pulses
from the resonance frequency $\omega_{0}$. In a frame rotating at
$\omega_{c}$, the transfer function reads
\begin{equation}
H\left(\ell,\Omega\right)=\exp\left[-\frac{L}{1+i\left(\Omega+\Delta\right)/\gamma}\right]\label{eq:trenteetun}
\end{equation}
where $\Delta$ is the detuning. When $e(0,t)=\widehat{e}_{2}(0,t)$,
we get for $\widehat{e}_{2}(\ell,t)$ the analytical expression
\begin{multline}
\widehat{e}_{2}(\ell,t)=\frac{3H\left(\ell,0\right)}{8}+\frac{H\left(\ell,\Omega_{2}\right)}{4}\exp\left(i\Omega_{2}t\right)\\
+\frac{H\left(\ell,-\Omega_{2}\right)}{4}\exp\left(-i\Omega_{2}t\right)+\frac{H\left(\ell,2\Omega_{2}\right)}{16}\exp\left(2i\Omega_{2}t\right)\\
+\frac{H\left(\ell,-2\Omega_{2}\right)}{16}\exp\left(-2i\Omega_{2}t\right)\label{eq:trentedeux}
\end{multline}
which is reduced to Eq.(\ref{eq:vingtneuf}) in the resonant case.
The envelope $\widehat{e}_{2}(\ell,t)$ is now complex and the corresponding
intensity $\left|\widehat{e}_{2}(\ell,t)\right|^{2}$ does not fall
to zero in its main part. 
\begin{figure}
\centering{} \includegraphics[width=0.95\columnwidth]{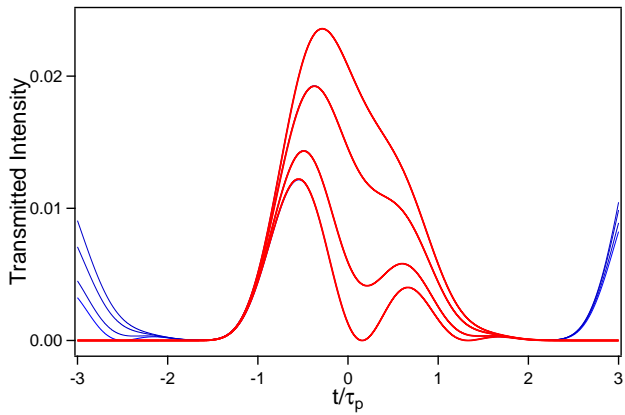} \caption{Effect on the transmitted intensity of a frequency detuning of the
incident pulse for $e(0,t)=e_{2}(0,t)$. Parameters $L=3$, ($C=26$
dB), $\tau_{p}\approx2.55/\gamma$ (as in Fig \ref{fig:Figure7})
and, from top to bottom, $\Delta/\gamma=\pm0.6$, $\pm0.5$, $\pm0.3$
and $0$. The solid red lines (the thin blue lines) are the exact
solutions (the analytical periodic solutions).\label{fig:Figure10}}
\end{figure}
This behavior is illustrated on Fig.\ref{fig:Figure10} obtained in
the conditions of Fig.\ref{fig:Figure7} for $\Delta/\gamma=\pm0.6$,
$\pm0.5$, $\pm0.3$ and $0$ (from top to bottom). As expected, the
advance (the pulse amplitude) decreases (increases) as a function
of the detuning, the fractional advance falling from $55\,\%$ for
$\Delta=0$ to $29\,\%$ for $\Delta=\pm0.6\gamma$. Note also the
disappearance of the secondary lobes and the \emph{strong asymmetry}
of the transmitted pulse when the detuning exceeds $0.4\gamma$. As
expected the analytical periodic solution $\left|\widehat{e}_{2}(\ell,t)\right|^{2}$
perfectly fits the response $\left|e_{2}(\ell,t)\right|^{2}$ to the
pulse of envelope $e_{2}(0,t)$ in the whole domain where $e_{2}(0,t)\neq0$
($\left|t\right|\leq\pi/\Omega_{2}\approx1.9\tau_{p}$). Similar results
are obtained when $C=50\,dB$ . Taking $\Delta=\gamma/5$ and $\tau_{p}=6/\gamma$,
it is in particular possible to reproduce the fractional advance ($F=68\,\%$)
and the shape (including its asymmetry) of the probe pulse which have
been recently evidenced in a four-wave mixing experiment \cite{sw17}.
Again the intensity profile is very well fitted by the analytical
periodic solution. We incidentally mention that, for the same contrast
and the same detuning, slow light is obtained for shorter durations
of the incident pulse. Otherwise said, a small detuning accelerates
the transition from fast to slow light. This phenomenon, easily explained
by a spectral analysis, is general and an asymmetry of the absorption
line has a comparable effect \cite{ta01,na09}.

Results obtained on resonance with a symmetric absorption line can
be extended to the case where the absorption is exactly compensated
by a gain which is also on resonance at the carrier frequency of the
incident pulses. The medium is then\emph{ transparent} at this frequency
and the transfer function of this twofold resonant medium reads
\begin{equation}
H\left(\ell,\Omega\right)=\exp\left(\frac{L}{1+i\beta\Omega/\gamma}-\frac{L}{1+i\Omega/\gamma}\right)\label{eq:trentetrois}
\end{equation}
where $L$ and $\gamma$ refer to the absorption line as previously
and $\beta$ is the ratio of the widths of the absorption and gain
lines ($0<\beta<1$). The transmittance $\left|H\left(\ell,\Omega\right)\right|^{2}$
actually equals $1$ for $\Omega=0$ (transparent medium). It is maximum
for $\Omega=\pm\gamma/\sqrt{\beta}$, yielding an intensity contrast
\begin{equation}
C=\exp\left[2L\left(\frac{1-\beta}{1+\beta}\right)\right].\label{eq:trentequatre}
\end{equation}
 For arbitrary $\beta$, the phase \cite{pa87} and the group advance
read
\begin{equation}
\Phi\left(\ell,\Omega\right)=\frac{L\Omega/\gamma}{1+\Omega^{2}/\gamma^{2}}-\frac{\beta L\Omega/\gamma}{1+\beta^{2}\Omega^{2}/\gamma^{2}}\label{eq:trentecinq}
\end{equation}
\begin{equation}
a_{g}=\left(1-\beta\right)\frac{L}{\gamma}=\left(1+\beta\right)\left(\frac{\ln C}{2\gamma}\right)\label{eq:trentesix}
\end{equation}
In the limit $\beta\rightarrow0$ (infinitely broad gain-profile),
these expressions are obviously identical to those obtained with a
single absorption line. The second form of $a_{g}$ in Eq.(\ref{eq:trentesix})
might lead to expect that, for a same contrast $C$, the systems with
$\beta\neq0$ are more efficient that the reference medium. This is
true for the absolute group advance but not for the fractional advance.
An examination of the phase $\Phi\left(\ell,\Omega\right)$ as a function
of $\Omega$ \cite{pa87} yields a qualitative explanation of this
apparent paradox. 
\begin{figure}
\centering{} \includegraphics[width=0.95\columnwidth]{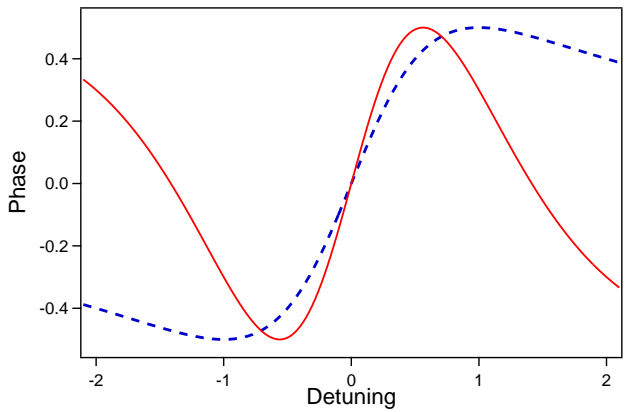}
\caption{Phase of transparent resonant media as a function of the detuning.
The solid red line (the dashed blue line) is obtained for $\beta=1/2$
($\beta=0$). The phase (the detuning) is expressed in units of $\ln C$
($\gamma$).\label{fig:Figure11}}
\end{figure}
Figure \ref{fig:Figure11}, obtained for a same value of $C$ when
$\beta=0$ and $\beta=1/2$, shows that the slope $d\Phi/d\Omega$
at $\Omega=0$ is actually larger for $\beta\neq0$ . On the other
hand the spectral domain where this slope is positive is narrower
and, for a same distortion of the transmitted pulse, the incident
pulse should be lengthened to the prejudice of the fractional advance.
In fact the maximum fractional advance does not significantly depend
on $\beta$ and is even independent of it in the range of validity
of the approximation $e\left(\ell,t\right)\approx\widehat{e}_{1}\left(\ell,t\right)$.
We then get
\begin{equation}
F\approx\frac{\Phi\left(\ell,\Omega_{1}\right)}{\Omega_{1}\tau_{p}}\approx0.44L\left(\frac{\Omega_{1}/\gamma}{1+\Omega_{1}^{2}/\gamma^{2}}-\frac{\beta\Omega_{1}/\gamma}{1+\beta^{2}\Omega_{1}^{2}/\gamma^{2}}\right).\label{eq:trentesept}
\end{equation}
$F$ is maximum for $2\beta^{2}\Omega_{1}^{2}/\gamma^{2}=\beta^{2}+4\beta+1-\left(\beta+1\right)\sqrt{\beta^{2}+6\beta+1}$.
This maximum reads
\begin{equation}
F_{m}\approx0.22L\left(\frac{1-\beta}{1+\beta}\right)=0.11\ln C\label{eq:trentehuit}
\end{equation}
 and is actually independent of $\beta$ for a given contrast. Moreover
a simulation made for $L=3$ and $\beta=1/2$ ($\ln C=2$) leads to
a normalized intensity profile that perfectly fits that obtained with
a single absorption line for a same contrast, that is for $L=1$ (Fig.\ref{fig:Figure6}). 

\section{DISCUSION\label{sec:DISCUSION}}

In the numerous theoretical articles on fast light published in the
last eighteen years, fast light in absorbing media is generally discarded
in favor of gain-assisted fast light (renamed superluminal light).
We point out on the contrary that\emph{, for a same fractional advance
and a same distortion }of the transmitted pulse, absorbing media are
preferable to media with gain. In the most cited experimental report
on fast light \cite{wa00}, the authors exploit the anomalous dispersion
associated with a minimum of gain between two gain lines created by
Raman pumping. As remarked in a premonitory article \cite{ste94},
the use of an amplifying medium raises some difficulties due in particular
to the phenomenon of amplified spontaneous emission and a risk of
lasing on the gain lines. Cross modulation instability also limits
the gain that can be used \cite{ste03b}. Another point is that, for
a given contrast, the gain-doublet arrangement yields weaker pulse
advances than those obtained with a single absorption line \cite{ma03,ma05}.
The fractional advance evidenced in \cite{wa00} is actually very
small (below $3\,\%$) and, as noted in \cite{ri02}, the advances
at the rise and at the fall of the pulse are significantly different
with a narrowing indicator $\xi=20\,\%$. Moreover it should be remarked
that, contrary to the title given in a subsequent article \cite{do01},
the medium is far from being transparent with a transmittance of only
$40\,\%$ and that the experimental results reported in \cite{wa00,do01}
can be reproduced with the single absorption-line arrangement with
a transmittance better than $66\,\%$. See Fig.1 in \cite{ma03}.
More convincing experiments using the gain-doublet arrangement are
reported in \cite{ste03a}. The effects of the cross modulation instability
\cite{ste03b} are here overcome by using two separate spatial regions
for the two Raman pump beams \cite{re4}. The fractional advance is
now $F\approx10\,\%$ with a narrowing indicator $\xi\approx13\,\%$.
We however remark that the same fractional advance can be obtained
in the single absorption-line arrangement with a fairly good transmittance
($14\,\%$) and a weaker pulse narrowing ($\xi\approx8\,\%$). See
Fig.\ref{fig:Figure4}. 

Independently of the abovementioned reasons that restrict the use
of gain media and of eventual problems of detection in the case of
absorbing media, one may wonder why fractional advances exceeding
$40\,\%$ \emph{with moderate distortion} have not been evidenced
in linear media. The response is obviously that the large contrasts
$C$ required to attain this purpose entail a dramatic sensitivity
of the transmitted pulse to defects of the incident one. As evidenced
in the experiment reported in \cite{se85} with $C\approx45\,dB$,
imperceptible quasi-discontinuities of the envelope of the incident
pulse or of its derivatives originate transients (wiggles) in the
envelope of the transmitted pulse which become invading when $C$
is enhanced (even slightly). Simulations also show that the effect
of a chirping of the incident pulse may become dramatic for large
contrast. Although this point is beyond the scope of the present article
(it is a subject \emph{per se}), it is also probable that the effect
of a partial light-incoherence, moderate for small contrast \cite{wa02},
will be destructive when this contrast increases.

Compared to other arrangements used to evidence fast light, the single
absorption-line arrangement has obviously the advantage of its simplicity,
from both experimental and theoretical viewpoints. We however mention
that the contrast required to attain a given fractional delay can
be slightly reduced by using a doublet of absorption lines \cite{ma03},
the experimental implementation of which seems quite feasible. When
the lines are separated by $2\gamma/\sqrt{3}$, the lowest order contribution
to the pulse distortion cancels and the medium transmittance is flattened
around $\Omega=0$. We then get $\ln C=3L$ and
\begin{equation}
\Phi\left(\ell,\Omega\right)=\frac{2L\left(\Omega/\gamma\right)\left(2/3+\Omega^{2}/\gamma^{2}\right)}{\left(4/3+\Omega^{2}/\gamma^{2}\right)^{2}-4\Omega^{2}/3\gamma^{2}}.\label{eq:trenteneuf}
\end{equation}
In the range of validity of the approximation $e\left(\ell,t\right)\approx\widehat{e}_{1}\left(\ell,t\right)$,
the fractional advance reads $F\approx\Phi\left(\ell,\Omega_{1}\right)/\left(\Omega_{1}\tau_{p}\right)\approx0.44\Phi\left(\ell,\Omega_{1}\right)$
and attains its maximum $F_{m}\approx0.39L$ for $\Omega_{1}\approx1.39\gamma$
($\tau_{p}\approx1.65/\gamma$). We have thus $F_{m}\approx0.13\,\ln C$
instead of $0.11\,\ln C$ with a single absorption-line. This means
in particular that the fractional delay obtained for $C\approx8.7\,dB$
with a single absorption-line (Fig.\ref{fig:Figure6}) would be obtained
for $C\approx7.3\,dB$ with a doublet. A simulation shows that the
corresponding intensity profiles are very close, both being perfectly
fitted by the analytical periodic solution of Eq.(\ref{eq:vingttrois}).
Similar results are obtained for large contrasts. An intensity profile
comparable to that obtained Fig.\ref{fig:Figure9} for $C\approx50\,dB$
with a single absorption-line can be obtained for $C\approx47\,dB$
, both profiles being now well fitted by the analytical periodic solution
of Eq.(\ref{eq:vingtneuf}). 

\section{CONCLUSION\label{sec:CONCLUSION} }

The observation of significant fast-light effects requires the use
of systems with a large contrast $C$ between maximum and minimum
transmissions. We have studied in detail the reference case of a dilute
medium with a narrow absorption line, the frequency of which coincides
with the carrier frequency of the incident pulse. The impulse response
$h\left(\ell,t\right)$ relating the envelope $e\left(\ell,t\right)$
of the transmitted pulse to that $e\left(0,t\right)$ of the incident
pulse is then real. In a retarded time picture (time delayed by the
transit time in vacuum), the group advance $a_{g}$ is positive no
matter the propagation distance and can be identified to the advance
of the center-of-gravity of $e\left(\ell,t\right)$ over that of $e\left(0,t\right)$.
This advance generally differs from the advance $a$ of the pulse
maximum and, quite generally, \emph{a large group advance is not
a sufficient condition to observe significant fast light effects}.

By convoluting $h\left(\ell,t\right)$ and $e\left(0,t\right)$, fast
light appears as resulting from the interference of the incident wave
as if it had propagated in vacuum with the wave reemitted by the medium.
Explicit analytical expressions of the convolution product are obtained
in the short and long pulse limits. In the first case, we get $a\ll a_{g}$
with moderate attenuation of the peak intensity. The experimental
evidence of this behaviour seems to be an open challenge. In the second
case, $a$ tends to $a_{g}$ and the peak intensity is reduced by
the factor $C$. Most experimental results have been obtained in conditions
more or less approaching these ones. In both cases the fractional
advance $F=a/\tau_{p}$ is not optimum. 

The observation of significant fractional advances leads to use incident
pulses of intermediate duration.\textbf{ }Analytical results are then
obtained by a method of periodic continuation of the incident pulses.
When the contrast $C$ is below $20\,dB$ as in most experiments actually
performed in optics, the duration $\tau_{p}$ maximising $F$ has
an explicit form and the corresponding advance $a$ is half of the
group advance $a_{g}$. This latter result remains a good approximation
when $C$ is increased but the main transmitted pulse is then followed
by large secondary pulses. Their lowering demands a lengthening of
the pulses that reduces $F$ below its maximum value.

The range of validity of the method of periodic continuation is very
broad. It enables one to reproduce analytically the asymmetry of the
transmitted pulses when the carrier frequency of the incident pulses
is detuned from resonance. In the case of a resonant transparent medium,
it shows that, for a same contrast, the maximum fractional advance
is roughly equal to that of our reference medium.

Insofar as the fractional advance is mainly determined by the transmission
contrast, the use of amplifying media seems unsuitable to evidence
significant fast-light effects. Large gains indeed originate optical
instabilities, obviously absent in absorbing media. Anyway, the use
of large contrasts requires an exceptional quality of the incident
pulse, the smallest imperfections of which can lead to considerable
distortion of the transmitted pulse. That is probably the main practical
limitation to fast light. Improvement of the quality of the incident
pulses combined with the use of a doublet of absorption lines instead
of a single line appears to be one way to evidence fast-light effects
beyond those observed up to now. 

\section*{{ACKNOWLEDGEMENTS}}

This work has been partially supported by the Minist\`{e}re de l'Enseignement
Sup\'{e}rieur, de la Recherche et de l'Innovation, the Conseil R\'{e}gional
des Hauts de France and the European Regional Development Fund (ERDF)
through the Contrat de Projets \'{E}tat-R\'{e}gion (CPER) 2015\textendash 2020,
as well as by the Agence Nationale de la Recherche through the LABEX
CEMPI project (ANR-11-LABX-0007).


\begin{thebibliography}{0}
\expandafter\ifx\csname natexlab\endcsname\relax\def\natexlab#1{#1}\fi
\expandafter\ifx\csname bibnamefont\endcsname\relax
  \def\bibnamefont#1{#1}\fi
\expandafter\ifx\csname bibfnamefont\endcsname\relax
  \def\bibfnamefont#1{#1}\fi
\expandafter\ifx\csname citenamefont\endcsname\relax
  \def\citenamefont#1{#1}\fi
\expandafter\ifx\csname url\endcsname\relax
  \def\url#1{\texttt{#1}}\fi
\expandafter\ifx\csname urlprefix\endcsname\relax\def\urlprefix{URL }\fi
\providecommand{\bibinfo}[2]{#2}
\providecommand{\eprint}[2][]{\url{#2}}

\end{thebibliography}


\begin{thebibliography}{10}
\bibitem{chi97} R.Y. Chiao and A.M. Steinberg, Tunnelling times and
superluminality, Prog. Opt. \textbf{37}, 345 (1997).

\bibitem{bo02} R.W. Boyd and D.J. Gauthier, \textquotedblleft Slow\textquotedblright{}
and \textquotedblleft Fast light\textquotedblright , Prog. Opt. \textbf{43},
497 (2002) 

\bibitem{to14} M Tomita, H. Amano, S. Masegi, and A.I. Takluder,
Direct Observation of a Pulse Peak Using a Peak-Removed Gaussian Optical
Pulse in a Superluminal Medium, Phys. Rev. Lett. \textbf{112}, 093903
(2014). 

\bibitem{ma05} B. Macke, B. S\'{e}gard, and F. Wielonsky, Optimal superluminal
systems, Phys. Rev. E \textbf{72}, 035601(R) (2005).

\bibitem{pa87} We use the definitions, sign convention and results
of the linear system theory. See, for example, A. Papoulis, \emph{The
Fourier integral and its applications }(Mc Graw Hill, New York, 1987).
Note that a different sign convention is often used in optics. The
passage from one convention to the other is made by replacing the
complex quantities by their conjugates. \emph{The phases are then
changed in their opposites }but the final results obviously do not
depend on the used convention.

\bibitem{wa02a} L.J. Wang, Causal \textquotedblleft all-pass\textquotedblright{}
filters and Kramers-Kronig relations, Opt. Commun. \textbf{213}, 27
(2002). 

\bibitem{ma16a} B. Macke and B. S\'{e}gard, Comment on: Gain-assisted
superluminal light propagation through a Bose-Einstein condensate
cavity system, Eur. Phys. J. D \textbf{70}, 193 (2016).

\bibitem{so03} D.R. Solli, C.F. McCormik, C. Ropers, J.J. Morehead,
R.Y. Chiao, and J.M. Hickmann, Demonstration of Superluminal Effects
in an Absorptionless, Nonreflective System, Phys. Rev. Lett. \textbf{91},
143906 (2003).

\bibitem{bru04} N. Brunner, V. Scarani, M. Wegmüller, M. Legr\'{e}, and
N. Gisin, Direct Measurement of Superluminal Group Velocity and Signal
Velocity in an Optical Fiber, Phys. Rev. Lett. \textbf{93}, 203902
(2004). 

\bibitem{ma16b} B. Macke and B. S\'{e}gard, Simultaneous slow and fast
light involving the Faraday effect, Phys. Rev. A \textbf{94}, 043801
(2016). 

\bibitem{ma03} B. Macke and B. S\'{e}gard, Propagation of light-pulses
at a negative group-velocity, Eur. Phys. J. D \textbf{23}, 125 (2003).

\bibitem{chu82} S. Chu and S. Wong, Linear pulse propagation in an
absorbing medium, Phys. Rev. Lett. \textbf{48}, 738 (1982).

\bibitem{se85} B. S\'{e}gard and B. Macke, Observation of negative velocity
pulse propagation, Phys. Lett. \textbf{109A}, 213 (1985).

\bibitem{ta03} H. Tanaka, H. Niwa, K. Hayami, S. Furue, K. Nakayama,
T. Kohmoto, M. Kunitomo, and Y. Fukuda, Propagation of optical pulses
in a resonantly absorbing medium: Observation of negative velocity
in Rb vapour, Phys. Rev. A \textbf{68}, 053801 (2003).

\bibitem{ke12} J. Keaveney, I.G. Hughes, A. Sargsyan, and C.S. Adams,
Maximal refraction and superluminal propagation in a gaseous nanolayer,
Phys. Rev. Lett. \textbf{109}, 233001 (2012).

\bibitem{je16} S Jennewein, Y.R.P. Sortais, J.F. Greffet, and A.
Browaeys, Propagation of light through small clouds of cold interacting
atoms, Phys. Rev. A \textbf{94}, 053828 (2016).

\bibitem{ak02} A.M. Akulshin, A. Cimmino, and G.I. Opat, Negative
group velocity of a light pulse in cesium vapour, Quantum Optics \textbf{32},
567 (2002).

\bibitem{go02} A. Godone, F. Levi, and S. Micalizio, Slow light and
superluminality in the coherent population trapping maser, Phys. Rev.
A \textbf{66}, 043804 (2002).

\bibitem{ki03} K. Kim, H.S. Moon, C. Lee, S.K. Kim, and J.B. Kim,
Observation of arbitrary group velocities of light from superluminal
to subluminal on a single atomic transition line, Phys. Rev. A \textbf{68},
013810 (2003).

\bibitem{ka04} H. Kang, G. Hernandez, and Y. Zhu, Superluminal and
slow light propagation in cold atoms, Phys. Rev. A \textbf{70}, 011801(R)
(2004).

\bibitem{mi04} E.E. Mikhailov, V. Sautenkov, I. Novikova, and G.R.
Welch, Large negative and positive delay of optical pulses in coherently
prepared dense Rb vapor with buffer gas, Phys. Rev. A \textbf{69},
063808 (2004).

\bibitem{br08} W.G.A. Brown, R. McLean, A. Sidorov, P. Hannaford,
and A. Akulshin, Anomalous dispersion and negative group velocity
in a coherence-free cold atomic medium, J. Opt. Soc. Am. B, \textbf{25},
C82 (2008).

\bibitem{ak10} A.M. Akulshin and R.J. McLean, Fast light in atomic
media, J. Opt. \textbf{12}, 104001 (2010).

\bibitem{wa00} L.J. Wang, A. Kuzmich, and A. Dogariu, Gain-assisted
superluminal light propagation, Nature (London) \textbf{406}, 277
(2000). 

\bibitem{ste03a} M.D. Stenner, D.J. Gauthier, and M. Neifeld, The
speed of light in a \textquoteleft fast-light\textquoteright{} optical
medium, Nature (London) \textbf{425}, 695 (2003).

\bibitem{chi07} S. Chi, M. Gonzalez-Herraez, and L. Thevenaz, Simple
technique to achieve fast light in gain regime using Brillouin scattering,
Opt. Express \textbf{15}, 10814 (2007).

\bibitem{gla12} R.T. Glasser, U. Vogl, and P.D. Lett, Stimulated
generation of superluminal light pulses via four-wave mixing, Phys.
Rev. Lett. \textbf{108}, 173902 (2012).

\bibitem{sw17} J.D Swaim and R.T. Glasser, Faster light with competing
absorption and gain, Opt. Express \textbf{26}, 10643 (2018).

\bibitem{ste94} A.M. Steinberg and R.Y Chiao, Dispersionless, highly
superluminal propagation in a medium with a gain doublet, Phys. Rev.
A \textbf{49}, 2071 (1994).

\bibitem{ste03b} M.D. Stenner and D.J. Gauthier, Pump-beam-instability
limits to Raman-gain doublet \textquotedblleft fast light\textquotedblright{}
pulse propagation, Phys. Rev. A \textbf{67}, 063801 (2003).

\bibitem{cri70} M.D. Crisp, Propagation of Small-Area Pulses of Coherent
Light through a Resonant Medium, Phys. Rev. A \textbf{1}, 1604 (1970).

\bibitem{re1} This result is easily retrieved from the transfer function
by means of Laplace transform procedure.

\bibitem{fe63} R. Feynman, R. Leighton, and M. Sands, \emph{The Feynman
Lectures on Physics}, \emph{Volume I} (Addison-Wesley, Reading 1963),
Ch.31.

\bibitem{re2} See footnote 10 in \cite{chi97}

\bibitem{bu04} N.S. Bukhman, On the relation between retarded and
advanced arrival of a pulse at the output of an optical system, Opt.
Spectrosc. \textbf{96}, 626 (2004). 

\bibitem{ma08} B. Macke and B. S\'{e}gard, Two-pulse interference and
superluminality, Opt. Commun. \textbf{281}, 12 (2008).

\bibitem{cha09} P. Chamarro-Posada and F.J. Frail-Pelaez, Superluminal
propagation in resonant dissipative media, Opt. Commun. \textbf{282},
1095 (2009).

\bibitem{do03} J.C.I. Dooges and J.P. O\textquoteright Kane, in\emph{
Deterministic Methods in Systems Hydrology} (CRC Press, Boca Raton,
2003), p.52.

\bibitem{pe00} J. Peatross, S.A. Glasgow, and M. Ware, Average Energy
Flow of Optical Pulses in Dispersive Media, Phys. Rev. Lett. \textbf{84},
2370 (2000).

\bibitem{ta01} A.I. Talukder, Y. Amagishi, and M Tomita; Superluminal
to Subluminal Transition in the Pulse propagation in a Resonantly
Absorbing Medium, Phys. Rev. Lett. \textbf{86}, 3546 (2001).

\bibitem{ta05} A.I. Talukder, T. Haruta, and M Tomita, Measurement
of Net Group and Reshaping Delays for Optical Pulses in Dispersive
Media, Phys. Rev. Lett. \textbf{94}, 223901 (2005{]}.

\bibitem{ko05} T. Kohmoto, H. Tanaka, S. Furue, K. Nakayama, M. Kunitomo,
and Y. Fukuda, Nonadherence to the conventional group velocity for
nanosecond light pulses in Rb vapour, Phys. Rev. A \textbf{72}, 025802
(2005).

\bibitem{na09} L. Nanda, H. Wanare, and S. Anantha Ramakrishna, Why
do superluminal pulses become subluminal once they go far enough?,
Phys. Rev. A \textbf{79}, 041806(R) (2009).

\bibitem{re3} Whatever $L$ is, $\left[\gamma\left(1+L/2\right)\right]^{-1}$
is a good estimate of the duration of $k\left(\ell,t\right)u_{H}\left(t\right)$
at $1/e$ of its maximum amplitude.

\bibitem{ma12} B. Macke and B. S\'{e}gard, Simple asymptotic forms for
Sommerfeld and Brillouin precursors, Phys. Rev. A \textbf{86}, 013837
(2012). 

\bibitem{ri02} H.L. Ringmacher and L.R. Mead, Comment on \textquotedblleft Gain-assisted
Superluminal Light Propagation\textquotedblright , arXiv:physics/0209012
(2002). 

\bibitem{do01} A. Dogariu, A. Kuzmich, and L.J. Wang, Transparent
anomalous dispersion and superluminal light-pulse propagation at a
negative group delay, Phys. Rev. A \textbf{63}, 053806 (2001). 

\bibitem{re4} The same strategy of separation of the pumping regions
is used in the experiments on an optical fiber reported in \cite{chi07}
where the pumping is ensured by Brillouin instead of Raman effect. 

\bibitem{wa02} L.G. Wang, N.H. Liu, Q. Lin, and S.Y. Zhu, Effect
of coherence on the superluminal propagation of light pulses through
anomalously dispersive media with gain, Europhys. Lett. \textbf{60},
834 (2002) 
\end{thebibliography}
\end{document}